# TellTable Spreadsheet Audit:
# from technical possibility to operating prototype

John Nash, Andy Adler and Neil Smith
School of Management, University of Ottawa, Ottawa, K1N 6N5, Canada
nashjc@uottawa.ca

**ABSTRACT**

*At the 2003 EuSpRIG meeting, we presented a framework and software infrastructure to generate and analyse an audit trail for a spreadsheet file. This report describes the results of a pilot implementation of this software (now called TellTable; see www.telltable.com), along with developments in the server infrastructure and availability, extensions to other "Office Suite" files, integration of the audit tool into the server interface, and related developments, licensing and reports. We continue to seek collaborators and partners in what is primarily an open-source project with some shared-source components.*

## 1. BACKGROUND

At the 2003 EuSpRIG Conference in Dublin, we presented our first, "proof of concept" version of a spreadsheet audit trail capability. This showed that it was possible to use the change-recording infrastructure of modern spreadsheet processors to keep track of changes in the underlying spreadsheet file. That is, we were able to record the time and authorship of all changes in each cell of the spreadsheet, and to analyse the contents before and after the change. Furthermore, by permitting filtering of the reported changes, we could focus on particular types of cell changes. For example, we have had a particular interest in changes from formulas to static values, since this type of change would be used to alter a course mark, a return on investment, or the result of an assignment calculation.

The audit trail reporting tool was one part of our work. To prevent users of a spreadsheet file from tampering with both the data and its audit trail, we needed to secure the file in some way. Our choice was to store the files on a server and to run the spreadsheet processor there. Software to do this was the second part of our reported work. At the time of the 2003 Conference, this was working, but we had constructed all the components "by hand". There was no simple installer nor any tools to facilitate many of the server administrative tasks. There had also not been a proof of concept of the whole system nor a pilot implementation.

In building the two complementary parts of the audit trail system, our efforts were considerably simplified by the existence of the OpenOffice spreadsheet processor *calc*, and the structure of the OpenOffice suite files in general. (Brauer, circa 2004; Sun Microsystems, 2002). These files are stored as compressed text with XML tags, for which there are freely available parsers (part of JVM for J2SE release 1.4; Apache Software Foundation, 1999, 2000; Free Software Foundation, 1999). Furthermore, *calc* maintains its user interface configuration in XML format. We are thus able to present the user with a spreadsheet processing menu that has the unwanted controls removed, such as those for change-recording. By turning on change recording and preventing the user from turning it off, we ensure the audit trail information is kept.

We believe it is likely that a similar strategy could be used to develop an audit trail for Microsoft Excel, Corel Quattro or Lotus 1-2-3. However, as far as we are aware, the internal file structures of these products are not published, making analysis of the files for audit very difficult. Some workers appear to have been able to learn the file structure, possibly by reverse engineering. Since we do not wish to have to rely on continuing efforts in such a pursuit, to date we have decided to work only with OpenOffice.org files.

## 2. DEVELOPMENTS SINCE JULY 2003

The principal developments in our work since the 2003 Conference have been as follows.

### 2.1 Software Improvements

Significant improvements have been made to the TellTable software, both in terms of usability and security. A read-only mode has been implemented, which is primarily used for viewing of archived file versions. Improvements to the user interface have allowed easier access to archives, and the ability to view, download or audit previous file versions. Security enhancements ensure that the server runs over an encrypted HTTP communication protocol, and rolling session ID's have been implemented to prevent replay and roll-back attacks.

### 2.2 A Pilot Study

During the September to December 2003Trimester at the University of Ottawa, we conducted a pilot study to use TellTable instead of a spreadsheet on a local computer for recording student marks from two sets of courses involving multiple professors and teaching assistants (TAs), in both English and French (Adler and Nash, 2004).

The principal finding of this pilot study has been that the TAs found it relatively easy to use, even over a dial-up line. Since TellTable runs OpenOffice.org applications on a server, transmitting the "screen" via a web-browser using VNC (Virtual Network Computing) technology, the satisfactory operation over dial-up lines was a pleasant surprise. The main complaint, which we are addressing, is that we made the server "too secure". When one user has a file open, other users are locked out. Should the first user not log out properly, the file remains locked. We are addressing this issue on a technical level with an automatic logout on non-activity or other events that indicate the session is over. However, there are decisions concerning the definition of an "abandoned" session and whether to implement time limits on sessions that are socio-political in nature. Our goal is to provide technical tools that enable managers to implement locally established policies.

There were also some complaints of failure that were the result of bugs in our code. At the time of writing these have been largely overcome. A final complaint was the lack of a cut-and-paste feature to allow external information to be copied into our files. This is discussed below.

### 2.3 GTEC Exhibition

In early October 2003, we were invited to participate in the Open Source Lab of the Government Technical Exhibition (GTEC) in Ottawa (MediaLive International, 2003) This was the impetus for choosing a new name for our system.

### 2.4 A Project Name

The working name for our project was SSScan, short for SpreadSheet Scanner. Given the extensions to the original ideas and the near unpronounceable short form, GTEC prompted us to think of a new name. Several candidates were considered, but TellTable fitted well with our feelings about the project. Also, it seemed a good fit in a bilingual environment where "tableur" is the term for "spreadsheet" in French. The domain name was registered and a web-site was set up in quick succession. Figure 1 shows the logo. The web address is **www.telltable.com (**17:30 31/03/2004**).**

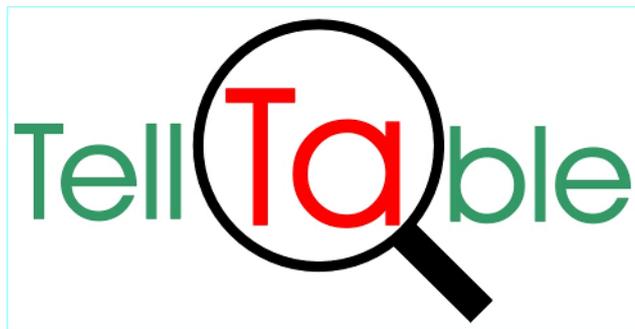

Figure 1. The TellTable logo.

**2.5 Other Office-Suite Applications**

During June of 2003, one of us (JN) had been working on some tools for Web Content Management. Another (AA) suggested talking to Dr. Sylvie Noël of the Communications Research Centre of Industry Canada, who was also working on tools for collaborative authoring of text-based documents. In mid-October 2003, recalling this discussion, Nash realized that the server structure could collaboratively operate non-spreadsheet functions of the OpenOffice.org suite. This led very rapidly to a joint paper submitted to a special issue of the IEEE Journal on Professional Communications (Adler, Nash, Noël, 2004). This paper, on software for collaborative authoring of documents, spreadsheets, presentations and drawings, was prepared using TellTable and thus a test of the ideas it discussed, in the process revealing some further strengths and weaknesses of the TellTable approach and implementation.

Again, dial-up access proved acceptable for viewing the file(s) or making small changes, though clearly less efficient than broadband access. The OpenOffice.org *writer* assigned different colours to the changes made by the different authors, making it very easy to see how we had edited the document.

On the negative side, the system does not properly allow cut-and-paste from existing files local to the user's machine into the server based objects, that is, text, spreadsheets, drawings or presentations. This problem is inherent in the use of a Java applet in an Internet browser-based client. The security model of the Java virtual machine does not allow cut and paste interaction with the local machine. This problem comes from the use of a web-browser interface, and is more general than our own software. However, it has particular implications for setting up documents or other files for use, for example, in providing the base class-list files for TAs to use for marks, or for resetting a file should an older version be needed. There are several possible approaches to solve this issue. One would be using a technology such as ActiveX to embed the applet into the Internet browser, which does not have the restrictions of the Java applet. Another approach, which we hope to implement shortly, allows files to be imported into a special paste window that is accessible to the browser screen. Although this is not quite the usual cut-and-paste protocol, it will allow material, such as bibliographic references, to be imported from existing files. Indeed it was the bibliographic reference task that brought home this deficiency.

So far, we have fairly extensive tests of the spreadsheet and word-processing functions, and have verified that slide presentations and drawings can be accessed. We will need "real" users to fully evaluate the latter two functions and any other features of OpenOffice that are or may become available.

## 2.6 The telltable-s Open Source Project

Early in January 2004, we established an open-source software project within the SourceForge framework to develop and enhance the server infrastructure. This project is called **telltable-s** (all lower case) and can be accessed via sourceforge.net, in particular at telltable-s.sf.net. The software can be downloaded using a CVS client (Cederqvist, 2002) and is licensed under the GNU Lesser General Public License. That is, anyone wishing to set up a TellTable server is at liberty to download the software and install it. There are many details in doing this that may make this tedious or difficult in some Linux/Unix distributions, and we anticipate some potential users will choose to engage us or other knowledgeable workers to help them. Furthermore, we hope that a growing community of users will encourage the development of tools to assist installation and maintenance of TellTable servers, as well as enlarge the pool of possible applications such servers can support.

In contrast, the TellTable Analyse spreadsheet audit trail analysis program is at this time considered proprietary software which we are distributing under a shared-source model. TellTable Analyse is *not* part of the **telltable-s** server project, though, as the next item describes, it can be conveniently run under the server infrastructure. Users who purchase a license for TellTable Analyse will get a copy of the source and run-time versions of the program, but may only use the program under agreed terms and conditions, and may not pass the program to others. We anticipate that there will be considerable interest in customized versions of TellTable Analyse, rather than a generic program. So far we have not begun developing audit trail tools for OpenOffice.org applications other than the spreadsheet processor.

While the primary feature of **telltable-s** is the server-based access to OpenOffice functions, there is no fundamental technical obstacle to running other software through the server, thereby allowing better control and management of important organizational files. While OpenOffice can read and write most files of type .doc, .xls, and .ppt, these are not the native format and we have some reservations about their suitability for audit. However, for situations where strict formatting of output is required, as in some desktop publishing requirements, it would be better to use the Microsoft Office tools. A technical objection to this – that Microsoft Office does not run under Linux – can be answered with the CodeWeavers CrossOver Office tools. We have performed a very cursory test that Microsoft Excel could be executed through the TellTable-s infrastructure, and could see no obvious failings. Whether Microsoft's EULA (End User License Agreement) permits or by legal action can be required to permit such a usage is still an open question. Nor are we certain how well we could control the change recording of Microsoft Office applications. Clearly it is of interest to us to resolve technical and legal issues so that TellTable can have the widest possible utility. We welcome an exchange of information with other workers, and EuSpRIG is clearly the premiere forum for such exchanges.

## 2.7 Workflow Tool Integration

Discussions with users and potential clients have reinforced our belief that TellTable capabilities should be well-integrated with workflow requirements. Recently we re-designed the user screens and incorporated a cleaner display where users can, according to their privileges, edit, view, audit or download the files to which they are permitted access. Some users may be granted only view or audit access, with no edit privileges. "View" presents the spreadsheet in the usual way, but does not allow changes to be made. "Audit" calls up the TellTable Analyse program that we described last year (Nash, Smith, and Adler, 2003). This provides ways to look at the detailed change record of the spreadsheet file, including filters to assist the user to focus on particular types of changes. As TellTable Analyse is written in Java, it can be easily run on the server or on the client machine. However, a single click implementation saves much of the work of downloading the file, launching TellTable Analyse, and finding and loading the correct version of the downloaded file.

We have also simplified and enhanced the version history interface, so that users with administrative privileges may download, view or audit any of the versions of a file. These priveleged users are able to insert files into the repository, or update a file with a version edited outside TellTable. While this means that such a user could remove the version history held within a spreadsheet file, TellTable maintains a read-only repository of all previous versions. As such administrative users are able to make arbitrary updates to the current file version, but a complete history of the modifications is still maintained.

TellTable Analyse has also been tested to show that it properly reports the effects of move or copy operations. OpenOffice *calc* keeps this audit information at the cell level, and visualization mechanisms to reconstruct the block operations are a feature we will have to keep in mind.

The single-click functionality can be enhanced by adding status and approval tags to files so that their disposition can be tracked and managed. This is very much in the spirit of management of the improvement of quality and productivity where we already have interests (Nash and Nash, 1997, 1998) under the rubric "Visible Management". The essential idea of Visible Management is to make visible the state of work processes. Clearly many workers are active in this area, as a Google search of "Business Processes" will quickly reveal. So far, our view is that it is easier to carry out office work processes than to make it clear that they have been done, and done correctly. We will be kept occupied for some time with incorporating such tools into TellTable.

**2.8 Applications in Education**

Two of us are academics who teach as well as carry out research. Therefore applications of a spreadsheet audit trail in educational environments were an obvious subject of interest, and we prepared a paper (Adler and Nash, 2004) for the electronic journal Spreadsheets in Education. The "how" in that paper is similar to the content of this paper, but the "what" and "why" are different.

**3. TELLTABLE DESIGN AND FUNCTION**

This section describes the design and architecture of TellTable, which has changed somewhat since 2003. TellTable can reside on one (or more) servers. Server functions are split into components running within the Apache web server and other processes running with limited privileges. For applications with many users and large spreadsheets, it is possible to spread the TellTable server components over several machines. Users connect to the server with a standard Internet browser which supports Java applets. Currently, TellTable only runs on Linux, but the architecture supports any UNIX platform. We have successfully tested TellTable user access with the Internet Explorer, Mozilla, Konqueror and Opera browsers running Windows 98/2000/XP, Linux, and Mac OS X, though we have noted that some versions do not give satisfactory operation.

The user types the URL of the server into the browser, or clicks on an appropriate pre-prepared link, and is presented with a standard login screen requiring a username and password. Once logged in, the user sees a list of files which he/she has permission to view or edit. At this point, the user is interacting with the web server components of TellTable; these manage the user login and the state of the user session, as well as spreadsheet file versions. Since web interactions themselves are stateless, user state is maintained using a randomly generated session ID in order to prevent out of sequence interactions with any TellTable application. For security the HTTPS web protocol is used.

Upon choosing to edit a file, the user sees a browser screen within which a Java applet presents a view of the spreadsheet running on the server. All keyboard and mouse activity in the browser is sent to the server to be processed, and the resulting spreadsheet view is sent back to the browser to be displayed. As indicated, we use the Virtual Network Computing (VNC) protocol to handle the display and interaction (Richardson, 2004). Some details follow.

When the server is started, a pool of limited privilege users (*suids*) are created and initialized to export their display using the VNC protocol, and to wait for information to arrive in an input directory. The VNC protocol allows a desktop display and keyboard and mouse activity to be exported over the Internet, using a low bandwidth protocol. Various clients have been developed to connect to a VNC exported display, including the Java applet used in TellTable. When a TellTable user selects a file to edit, the web server first verifies that a *suid* is available, and then copies the selected file, along with the information about the username and a randomly generated one time password into the input directory of the *suid*. The *suid* will then move the spreadsheet file to a processing directory, modify its configuration to match that of the given username, and open the file with the Openoffiec.org *calc* application into the desktop exported by VNC. In this way, we ensure that OpenOffice.org will record change information under the correct username. The web page sent to the user contains information to load the appropriate VNC viewer Java applet and to connect to the appropriate *suid*. When the user is finished editing the file and closes OpenOffice.org *calc*, the *suid* will move the edited file to its output directory, where it is further managed by the web server. For security, the VNC password is modified to a newly generated and encrypted password, while this password is also sent to the user's browser with the applet information so the file and user can be kept linked. We are aware that allowing network access to server user account creates potential security vulnerabilities. To address these issues, we limit the privileges of the *suid* in terms of its ability to access files on the server outside of its local directories.

After editing the file, the user has the option to save it or discard the changes, closes the Openoffice.org *calc* application, and then clicks on a web link in the browser page to return to the TellTable main menu. The edited file is tested for modifications, and is stored in the TellTable version repository. If a user wishes to view previous file versions, a version history page is available. Each previous version, including the name of the user who made the last edits and the time of saving is available to be viewed, audited, or downloaded. We do not permit editing of previous versions, as the semantics and workflow of such an edit are unclear. File versioning is managed using the CVS version control system. CVS software allows concurrent editing of files; however, such concurrent editing requires the ability to merge conflicting edits. As we are not aware of a robust approach to manage conflicts in spreadsheet files, we do not permit this within TellTable.

As indicated, the TellTable framework is flexible enough to allow diverse software to be run to perform different operations on spreadsheets and other office-suite files. For example, the TellTable Analyse spreadsheet audit software allows analysis of changes recorded in the spreadsheet files. We envisage other applications, both open-source and proprietary, being used within this framework as a way to enhance workflow management and efficiency.

## 4. PERFORMANCE TESTING

TellTable shifts much of the computational burden from the local machine to the server. In this section we report some tests to measure the relative loads of OpenOffice.org *calc* on the client in comparison to the TellTable server. Tests were performed using an Intel Pentium 4, 2.4 GHz server with 256 MB Ram running Mandrake Linux 9.1 within the VM Ware software environment. Client connections were made from an identical computer running windows XP using Internet Explorer with Sun Java 1.4.

### 4.1 Memory Test

The goal of this test was to measure the memory required by TellTable as a function of the number of users supported. The TellTable server was configured to support *N suid* processes. Initially, the memory requirements were measured without users editing spreadsheets. Subsequently, users were logged into all *suid* processes to edit a simple spreadsheet. The total "committed" server memory was measured for *N*= 2, 4 and 6, using the UNIX command "top". Measured memory consumption values were fitted to a line as:

$$151 + 5.8 \text{ (available } suid) + 17.0 \text{ (used } suid) \text{ MB}$$

This result indicates that the server memory requirements are relatively small compared to that required by the operating system and the spreadsheet data itself.

**4.2 Timing Test**

The goal of this test was to compare the time to carry out spreadsheet calculations within the TellTable framework versus similar calculations on a local machine. Tests were designed to perform a large computational load without excessive memory demands. The test spreadsheet is described in the appendix. For our tests we chose a spreadsheet size of 50x50. Results follow:

|         | *TellTable* | *Local machine (Win XP)* |
|---------|-------------|--------------------------|
| 1 user  | 81.5s       | 79.5s                    |
| 2 users | 165s        | N/A                      |

These results indicate a slight (2%) reduction in performance under TellTable, and another slight (1%) reduction in performance due to simultaneous use. However, these performance decreases appear to be acceptably small, especially given the benefit of an assured audit trail.

**5. WORK IN PROGRESS**

As the time of writing (end of March, 2004), we are testing the SourceForge version of telltable-s on a system running the Xandros Linux distribution (version 2.0) which is built upon the Debian distribution and package manager (www.debian.org 17:30 31/03/2004). Our pilot study ran on a system that used the Mandrake distribution (version 9.1), and there are a number of minor but annoying changes that need to be made to account for differences in file system layouts. We hope to build both documentation and scripted customization to allow the installation of telltable-s to be very simple. At the present time we do not have plans to port telltable-s to non-Linux servers, but we would provide what advice we can to anyone wishing to do so in exchange for learning what issues arise. In principle, such a port should not be difficult.

We are also working toward the development of a "live CD" demonstration tool based on the popular Knoppix version of Linux. Knoppix is also based upon Debian and there are a number of derivative systems. See www.knoppix.org (17:30 31/03/2004). The advantage of such CD-based tools is that they do not require installation to be tested and run, yet provide all the components of the operating system and supporting software that are needed, and provide them in a coherent and well-organized form. The user of most Intel or AMD PCs simply puts the CD in his/her computer and "reboots". After the session is over, the CD is ejected and the user restarts the machine to return to the installed operating system, which is not altered by the Knoppix software apart perhaps from a single "swapfile" to provide for virtual memory. (This file is only created with user permission, is well-identified and may be deleted afterwards.)

There are some limitations to the use of Java applets in the client browsers. While almost all modern browsers support Java applets, most do not install with the Java tools, but require a subsequent download. Examples include recent releases of Internet Explorer and Mozilla. Another difficulty is that the Java applet security model does not support access to the local machine clipboard, preventing the use of cut-and-paste from the local machine to TellTable. Since Internet Explorer has built-in support for ActiveX plugins, we are exploring the use of ActiveX applets for VNC. Fortunately, the VNC protocol is well supported, and others have built such support (e.g., http://www.veridicus.com/tummy/programming/vncx/ 23:20 08/04/2004). Modifications to the TellTable server are minor, involving modification to the web page that loads the VNC viewer applet. One remaining difficulty concerns testing (from the server) that the user's browser is capable of running various applet plugins.

## 6. CONCLUSION

What we now call the TellTable project only began in November 2002. Since that time we have managed a proof of concept of a spreadsheet audit trail system, run a pilot study, demonstrated and used a collaborative office suite and prepared a number of academic and professional papers.

This progress confirms our belief that the TellTable framework is a valuable approach to managing some of the risks in spreadsheet design. From the point of view of the enterprise, TellTable allows the server administrator to control the versions and audit history of spreadsheet files. This capability is critical to any organization because of the value of the information that is stored in spreadsheet form. In our own experience, the ability to quickly access previous file versions has become an indispensable tool. Furthermore, the TellTable approach avoids the problem of inflexible proprietary applications, usually built around database technology, by providing the flexibility of full spreadsheet capabilities. Anecdotal evidence suggests that rigid centralized applications are left largely unused because they don't match user requirements, while the "real" data is entered into spreadsheets, sometimes referred to as "black book" applications.

From both the individual and the organizational perspective, the capability of TellTable to provide an audit trail to spreadsheet changes and additions allows for the discovery and correction of errors or falsification of data, particularly when the audit information can be filtered and summarized by TellTable Analyse. This capability, which can be extended to other office-suite files, is believed to be unique at the time of writing.

## 7. ACKNOWLEDGEMENTS

We appreciate the support of Joseph Potvin and Marcel Boulianne of Public Works and Government Services Canada, and for inviting us to participate in GTEC 2003, and for help in correcting translations of TellTable information. Mary Nash helped edit this paper.

## 8. APPENDIX

This section describes the high-computational-load spreadsheet test. The tests in this paper are based on a spreadsheet that requires significant computations, without the need for very large spreadsheet files which would, instead, be a test of the physical and virtual memory systems.

Since we were unable to find such a test in the literature, a custom test spreadsheet was designed. We describe it here because it may constitute a useful benchmark. See the telltable-s distribution files
http://cvs.sourceforge.net/viewcvs.py/telltable-s/telltable-server/repository/files/timetest.sxc
(2004-6-4 15:30).

The test spreadsheets are built up with a border (left and top) that is two cells thick, and a main section which is cell C3 and all cells below and right of it. Cell B2 contains an arbitrary number, and recalculation is forced by changing its value. The "border" is built up of number increments and trigonometric functions such that the values are in a reasonable range to help ensure the matrix inverses described below are numerically stable. Each cell in the "main" section of the spreadsheet is the upper left or 1,1 element of the matrix inverse of the largest available square block of cells to the upper left of that cell. The test is then performed by adjusting the value of the arbitrary number in cell B2, and waiting for the spreadsheet to recalculate. The size of the spreadsheet is determined by the number of elements in column A and row 1 to the right and below cell C3. The spreadsheet structure is shown in the following table:

|   | A | B | C | D |
|---|---|---|---|---|
| 1 |   | 0 | =B1+1 | =C1+1 |
| 2 | 0 | *Arbitrary number* | =sin( B2+B1 ) | =sin( C2+C1 ) |
| 3 | =A2+1 | =cos( B2+A2 ) | *Main Section* | |
| 4 | =A3+1 | =cos( B3+A3 ) | | |

where element C3 in the *Main Section* is defined as

```
=INDEX(
  MINVERSE(
   OFFSET(
    $A$1;
    1 + MAX( 0; $A3 - C$1 );
    1 + MAX( 0; C$1 - $A3 );
    MIN( C$1;$A3 );
    MIN( C$1;$A3 )
   )
  );
 1;
 1)
```

This cell formula is then replicated for all cells below and right of C3. This spreadsheet has been tested on the spreadsheets OpenOffice.org *calc*, Microsoft Excel, and Gnumeric. Note that all semicolons (;) must be replaced by commas (,) for Excel and Gnumeric. We have not been able to make this spreadsheet work properly in Quattro Pro (version 9 was tried).

This spreadsheet was tested on a Pentium 4, 2.4 GHz computer running Windows XP for spreadsheets of size 30x30, 35x35, 40x40, 45x45, and 50x50, using OpenOffice.org *calc* version 1.1. Recalculation times for each size were 3.5, 7.5, 21, 44 and 80s, respectively. The maximum memory consumption of the spreadsheet software was 32.4, 33.2, 33.8, 34.4, and 35.3 MB, respectively. These data can be modelled as:

$$\text{recalculation time} = (\text{size} / 24.7)^{6.3} \text{ sec.}$$

and

$$\text{maximum memory} = 28.2 + 0.14 \text{ (size) MB}$$

Thus, this spreadsheet shows strong exponential growth in calculation time, for linear growth in memory consumption.

We have also built a moderately computationally intensive spreadsheet that allows us to test the time/memory relationship for very large spreadsheets. With this we can explore situations where the spreadsheet file cannot be stored entirely in memory, but must be swapped out to disk (virtual memory). This test is based on one in Nash and Nash (1994, p. 67). It uses a first row (Row 3) where cell A3 is set to 1, and cell B3 is set =A3+1, and so on to increment each element of the third row by 1. Cell A4 is set =A3+2, so that the first column increments each row by 2. We can make this border have as many rows and columns as we like, then fill in the rectangle with top-left cell B4 do the maximum settings with cell i, j computing the function

$$\exp(\sin(\cos(\text{rowdata} + \text{columndata})))$$

That is, cell E7 is set =EXP(SIN(COS($A7*E$3))). We can make the core function more or less complicated as needed for our tests, but must note that this spreadsheet is less time-consuming than the one above. It is available as http://cvs.sourceforge.net/viewcvs.py/telltable-s/telltable-server/repository/files/timetest.sxc (2004-6-4 15:30).

Our objective in building these tests was to see how badly the TellTable server would degrade when running more than one compute-bound spreadsheets. Very large spreadsheets may also run slowly due to the use of swap memory on disk, which is an artifact of virtual memory subsystems.

## 9. REFERENCES

Adler, A. and Nash, J. C. (2004) Knowing what was done: uses of a spreadsheet log file. Spreadsheets in Education, 1(2):118-130, http://www.sie.bond.edu.au/articles/1.2/AdlerNash.pdf

Adler, A., Nash, J.C. And Noël, S. (2004) Challenges in Collaborative Authoring Software, submitted to a special Issue of the IEEE Transactions on Professional Communication "Expanding the boundaries of E-collaboration".

Apache Software Foundation (2000) Crimson http://xml.apache.org/crimson/ 15:32 31/03/2004.

Apache Software Foundation (1999) Xerces Java Parser 1.4.4 Release http://xml.apache.org/xerces-j/index.html 15:34 31/03/2004

Brauer, M. (circa 2004) OpenOffice.org XML File Format (main project page) http://xml.openoffice.org 15:20 31/03/2004.

Cederqvist, Per (2002) Version management with CVS, Bristol UK: Network Theory

Free Software Foundation (1999) GNU Lesser General Public License version 2.1, http://www.gnu.org/copyleft/lesser.html 17:00 31/03/2004.

Free Software Foundation, Inc. (1999) The GNU JAXP Project http://www.gnu.org/software/classpathx/jaxp/ 15:30 31/03/2004.

MediaLive International, Inc. (2003) Semaine GTEC Week, E-Canada: Governments Working Together to Serve Canadians. (Open Source Lab) http://www.gtecweek.com/ottawa2003/english/event_features/index.php?s=exhibition_details 15:45 31/03/2004.

Nash, J. C. and Nash, M. M. (1994) Scientific Computing with PCs, Nash Information Services Inc.: Ottawa, November 1994, 199 pages. Republished as part of the 8-book set "Dr. Dobb's Essential Books on Numerics and Numerical Programming", San Mateo CA: Miller Freeman Inc., December 1998.

Nash, J. C. and Nash, M. M. (1997) The Visible Management system: management ideas with library principles, in Communication and Information in Context: Society, Technology and the Professions, (Bernd Frohmann, editor), Proceedings of the 25th Annual Conference, Canadian Association for Information Science: Toronto, pp. 124-130, June.

Nash, J. C. and Nash, M. M. (1998) Visible management for design, programming and other creative processes, Proceedings of the CSME Forum 1998, Vol. 3, (Editors M A Rosen, D Naylor, and J C Keewall), Ryerson Polytechnic University, pp. 224-230.

Nash, J. C., Smith, N. and Adler, A. (2003) Audit and change analysis of spreadsheets, Proceedings of the 2003 Conference of the European Spreadsheet Risks Interest Group, eds. David Chadwick and David Ward, Dublin, London: EuSpRIG, pp. 81-90. July.

Richardson, T. (2003)The RFB Protocol, Version 3.7, RealVNC Ltd, (12 August), available as file rfbproto.pdf from http://www.realvnc.com/documentation.html 00:15 09/04/2004.

Sun Microsystems, Inc. (2002) OpenOffice.org XML File Format 1.0 Technical Reference Manual, Version 2, December. Downloadable as http://xml.openoffice.org/xml_specification.pdf 15:25 31/03/2004.